\documentclass[twocolumn,prd,floatfix,nofootinbib,preprintnumbers]{revtex4}
\usepackage{graphicx}

\begin{document}

\preprint{NT@UW 02-38}
\title{Solution of coupled vertex and propagator Dyson-Schwinger
  equations in the scalar Munczek-Nemirovsky model} 
\author{W. Detmold} 
\affiliation{Department of Physics, University of
  Washington, Box 351560, Seattle, WA 98195, U.S.A.}
\begin{abstract}
  In a scalar $\varphi^2\chi$ model, we exactly solve the vertex and
  $\varphi$ propagator Dyson-Schwinger equations under the assumption
  of a spatially constant (Munczek-Nemirovsky) propagator for the $\chi$
  field. Various truncation schemes are also considered.
\end{abstract}  
\maketitle

\section{Introduction}

The Dyson-Schwinger equations (DSEs) \cite{DSE} are an infinite system
of coupled integral equations for the Green functions of a quantum
field theory.  In principle, their solution contains all possible
information in the theory. However, the DSE for an $n$-point Green
function involves various other Green functions of order $m\leq n+2$.
So to make progress, the system must be truncated at some stage and
models must be made for the unknown Green functions.  For example, in
QCD, the DSE for the quark propagator involves not only the quark
propagator, but also the gluon propagator and the full quark-gluon
vertex, both of which need to be calculated or modelled in some way
for the equation to provide a solution for the quark propagator.

Notwithstanding the need for model input, the DSEs provide a
relativistically covariant, non-perturbative approach to field theory
that has enjoyed considerable success in recent years.  In QCD, models
based on the DSEs are able to describe accurately many facets of meson
and, more recently, baryon physics in free space and at nonzero
temperature and density \cite{Roberts:dr,Roberts:2000aa}.
Additionally, recent DSE studies are now providing insight into the
infrared behaviour of QCD Green functions and the nature of
confinement \cite{IRreview,gluonghost}. Sophisticated studies have
also been made of the DSEs in QED and other theories
\cite{Roberts:dr,Kizilersu:2000qd}.

Almost all DSE based studies to date focus on the equations for some
of the two point functions of the theory, making various models for
the remaining two- and three-point functions.  In this paper, we
examine a simple theory and solve a larger system of equations. We
consider a $\varphi^2\chi$ scalar field theory with a
Munczek-Nemirovsky (MN) model propagator for the $\chi$ particle
(which we introduce in Sec.~\ref{model}) and solve the coupled DSEs
for the $\varphi$ propagator and 3-point vertex. An exact solution is
possible in this model because relations amongst the Green functions
allow the remaining system of DSEs to close without truncation. We
also investigate the effect that imposing various truncations has on
the solutions.

The central result of this analysis is that there is no real solution
to the untruncated model unless the coupling is imaginary despite the
model having solutions for real couplings when additional truncations
of the DSEs are made. This is shown by the divergence of the
diagrammatic expansion of the vertex.

The rest of this paper is structured in the following manner.  We
introduce the details of the model in Sec. \ref{model}, and the
various truncations we employ in Sec. \ref{truncate}. In Sec.
\ref{fullsolution} we show that the DSEs of the Munczek-Nemirovsky
model close and derive their exact solution where it exists.  Finally,
we present our conclusions in Sec. \ref{conclusion}.

\section{Scalar model}
\label{model}

\begin{figure}
  \includegraphics[width=\columnwidth]{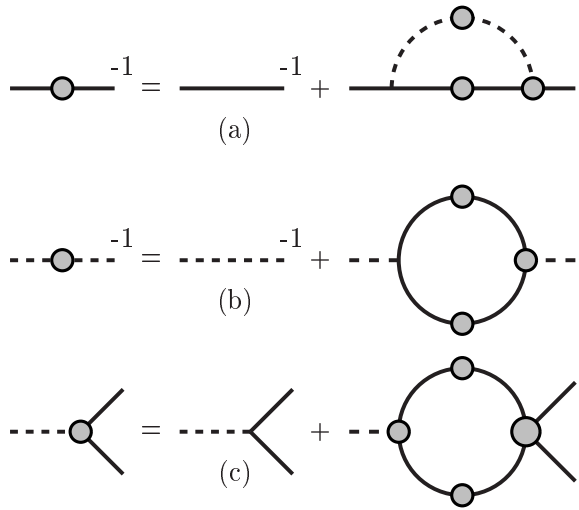}
  \caption{\label{fsde} The Dyson-Schwinger equations for the
    (a) $\varphi$ and (b) $\chi$ propagators and (c) vertex function
    of the scalar theory of Eq.~(\ref{lagrangian}). Solid (dashed)
    lines are $\varphi$ ($\chi$) propagators and grey circles denote
    full one particle irreducible components.}
\end{figure}

We consider the theory of two scalar fields, $\varphi$ and $\chi$,
described by the interaction Lagrangian:
\begin{equation}
\label{lagrangian}
{\cal L}_{\rm int} = g\ \varphi^2\chi \, ,
\end{equation}
and work in Euclidean space.  This scalar theory is primarily studied
as a precursor for an investigation in QCD, but is also interesting in
its own right and is sometimes used as a simplified description of
meson-nucleon interactions. We define the momentum- (and position-)
space propagators for the fields $\varphi$ and $\chi$ as $S(p)$
($S(x,y)$) and $D(p)$ ($D(x,y)$) respectively. These objects satisfy
the Dyson-Schwinger equations:
\begin{equation}
\label{phisde}
S^{-1}(p)= S_0^{-1}(p) +\Sigma(p)
\end{equation}
and
\begin{equation}
\label{chisde}
D^{-1}(p)= D^{-1}_0(p)+\Pi(p)\,,
\end{equation}
where the bare propagators are $S_0^{-1}(p)=(p^2 + m^2)$ for the
$\varphi$ field of mass $m$ and $D_0^{-1}=(p^2+M^2)$ for the $\chi$
field (mass $M$). The respective self-energies are defined as
\begin{equation}
  \label{SE}
  \Sigma(p)=\int_q g\  D(q) S(p-q) \tilde\Gamma (p-q,p)
\end{equation}
and
\begin{equation}
  \label{chiSE}
  \Pi(p)=\frac{1}{2}\int_q g\  S(q) S(p+q) \tilde\Gamma (q,p+q)\,,
\end{equation}
where $\tilde\Gamma(p,k)$ is the proper three-point vertex.  (Here and
in what follows momentum integrals are abbreviated as
$\int_q=\int\frac{d^4q}{(2\pi)^4}$.) These equations are shown
diagrammatically in Fig.~\ref{fsde}. An additional (constant) tadpole
contribution to Eq.~(\ref{SE}) is omitted as it can be absorbed into
the definition of the renormalised mass, $m$ \cite{Kuster:1996kx}.
Field renormalisation constants are suppressed as the ultraviolet
divergences of the theory will be eliminated by the specific form of
the $\chi$ propagator that we introduce below.

The dependence of these equations on the vertex function
$\tilde\Gamma$ means that they do not form a closed, computable
system.  An often used approach is to truncate the system at this
level by introducing a model for $\tilde\Gamma$ and solve the
resulting equations \cite{Ahlig:1998qf,Tjon,Sauli:2002qa}.\footnote{In
  phenomenological approaches to QCD, only the analogous equation for
  the quark propagator is retained and the combination of the gluon
  propagator and quark gluon vertex is
  modelled\cite{Roberts:2000aa}.}  We investigate various possible
truncations in Sec.~\ref{truncate}.

However, the vertex function satisfies its own DS equation [an
inhomogeneous Bethe-Salpeter equation---see Fig.~\ref{fsde}(c)] in
terms of the propagators and an additional unknown function, the
four-$\varphi$ scattering kernel, $K$:
\begin{eqnarray}
  \label{vertex}
  \Gamma(p,k)&\equiv& g^{-1}\tilde\Gamma(p,k)\\
  &&\hspace*{-13mm}=\ 1 +\int_q K(p,k,q) S(p+q) 
  \Gamma(p+q,k+q) S(k+q)\,. \nonumber 
\end{eqnarray}
Here we have rescaled $\tilde\Gamma\to\Gamma$ to absorb the bare
coupling of the theory. The scattering kernel, $K$, occurring in this
equation is two particle irreducible with respect to the $\varphi$
legs and contains no annihilation contributions \cite{Roberts:dr}. The
same kernel also enters in investigations of two-body bound states in
the Bethe-Salpeter approach \cite{Salpeter:sz}. By keeping
Eq.~(\ref{vertex}), it is possible to truncate the system at a higher
level, modelling $K$ rather than $\Gamma$. This automatically ensures
that the Dyson-Schwinger and bound state equations are truncated
consistently.

In this paper, we will assume that the propagator for the $\chi$ field
is constant throughout space-time. In momentum space this results in
\begin{equation}
\label{MN}
g^2 D(p) \equiv (2\pi)^4 \alpha \ \delta^4(p)\, ,
\end{equation}
where $\alpha$ is the (constant) strength of the correlation. This is
analogous to the approximation of Munczek and Nemirovsky (MN)
\cite{Munczek:1983dx} which has often been applied to model QCD
\cite{MNpapers,Bender:1996bb,Bender:2002as}. It was originally
motivated as a simple means with which to provide the necessary
infrared enhancement in the kernel of the quark DSE of QCD.  However
its physical foundations are fairly tenuous as it does not allow
momentum transfer between quarks or their consequent scattering.
Whilst it has this and other difficulties \cite{thesis}, it also has
the distinct advantage that the various integral equations of the DS
system reduce to algebraic equations and it remains an interesting
model.

With this scalar MN model in place, we will explore varying possible
truncations in the following section and then show that they are
unnecessary by solving the full set of equations and calculating the
exact solutions (where they exist) in Sec.~\ref{fullsolution}.

\section{Truncation schemes}
\label{truncate}

In this section, we shall consider a number of truncations that can be
applied to solve the DSEs above, each of which successively improves
on the last.  For simplicity, we omit the momentum dependence of
propagators and vertices in what follows since the MN model decouples
all differing momenta. In fact, the momentum only enters into the
propagator and vertex DSEs though the bare propagator,
$S_0^{-1}=p^2+m^2\equiv x$, so $\Gamma$, $\Sigma$, $S$ and $K$ depend
only on $x$.

It is possible to define the dimensionless quantities
$S^\prime=\sqrt{|\alpha|}S$, $\Sigma^\prime=\sqrt{|\alpha|}\Sigma$,
$\Gamma^\prime=\Gamma$, $x^\prime=x/\sqrt{|\alpha|}$ and
$K^\prime=K/|\alpha|$. If one works in terms of these new objects, all
dependence on the magnitude of $\alpha$ will factorise in our
analysis.  Consequently, only the sign of this parameter is important.
In order to avoid a proliferation of factors of $\alpha/|\alpha|$, we
shall use the original (unprimed) quantities and only consider
$\alpha=\pm1$.

\subsection{Rainbow truncation}
\begin{figure}
  \includegraphics[width=0.95\columnwidth]{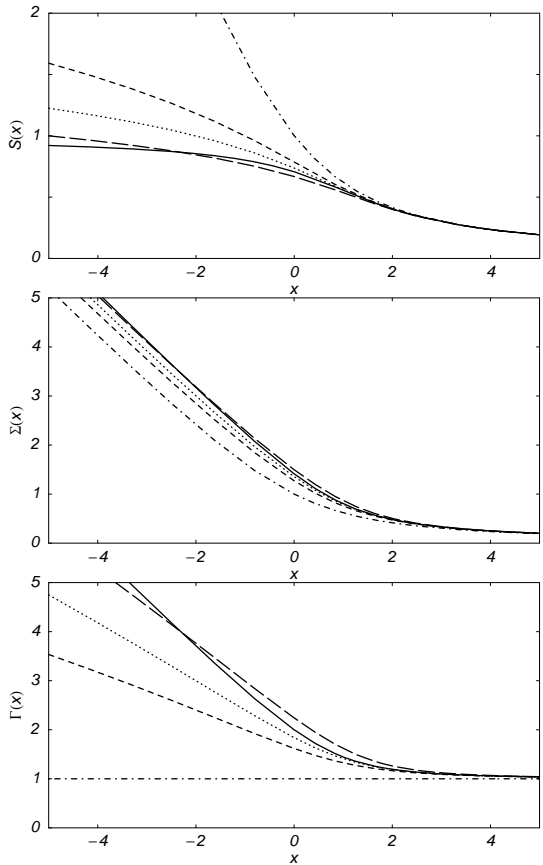}
  \caption{\label{fig1} Dependence of the propagator (upper
    panel), self-energy (middle panel) and vertex function (lower
    panel) on the variable $x=p^2+m^2$ for $\alpha=1$. Solutions are
    shown for the bare- (dot-dashed), the one-loop- (short-dashed),
    two-loop- (long-dashed), two-loop-ladder- (dotted) vertices and
    for the ladder truncation of the vertex DS equation (solid).}
\end{figure}

For this scalar theory, the rainbow truncation is defined by
\begin{equation}
  \label{rainbow}
  \Gamma_{\rm rainbow}(p,k)=1\,.
\end{equation}
In QCD, when the analogous truncation of the quark DSE is combined
with the ladder truncation of mesonic Bethe-Salpeter equations, the
resulting models successfully describe the pseudo-scalar and light
vector meson spectrum \cite{rainbowladder} with additional
higher-order corrections providing only minor modifications
\cite{Bender:1996bb,Bender:2002as}.  If the rainbow truncation is
combined with the Munczek-Nemirovsky propagator, Eq.~(\ref{MN}), the
DSE for the $\varphi$ propagator reduces to a simple quadratic
equation whose solutions are
\begin{equation}
  \label{baresoln}
  S(x)=\frac{x \pm {\sqrt{x^2 + 4\,\alpha }}}{2\,\alpha }\,.
\end{equation}
The physical solution\footnote{It is important to note that for each
  truncation considered here, the algebraic equations have only one
  solution that asymptotes to the bare propagator for large momentum
  or particle mass.}  and the corresponding self energy are shown as a
function of $x=p^2+m^2$ as the dot-dashed curve in Fig.~\ref{fig1} for
$\alpha=1$.

\subsection{One- and two-loop truncations}
In the algebraic MN model, it becomes simple to include any finite
order of contributions to the vertex. As no momentum flows along
$\chi$ propagators and there are no complicating Dirac structures,
every diagram at a given order contributes the same factor, For
example, the one- and two- loop vertices are given by
\begin{equation}
  \Gamma_{\rm 1\ loop} = 1 + \alpha\ S^2, \quad
  \Gamma_{\rm 2\ loop} = 1 +\alpha\ S^2 + 4\alpha^2 S^4\, ,
\end{equation}
respectively, where the factor of $4$ arises from diagram counting
(see Sec.~\ref{diagrams}).  These additional contributions to the
vertex are not perturbative corrections; they involve the full
propagator functions. Inserting these forms into the propagator DSE
leads to 4th- and 6th- order equations for S whose physical solutions
are also shown in Fig.~\ref{fig1} as the short- and long- dashed
lines, respectively. The lower panels of the figure show the
corresponding self energies and vertex functions. For comparison, we
also show the solution (dotted line) when only the ladder dressings of
the vertex are included to second order (that is, $\Gamma_{\rm 2\ 
  loop}$ with a factor of 1 instead of 4) which we label as
$\Gamma_{\rm 2\ loop}^\prime$.

\subsection{Vertex ladder truncation}
If (instead of explicitly approximating the vertex) we make a ladder
truncation of the vertex kernel, $K(p,k,q)=g^2 D(q)$, we are selecting
an infinite class of planar diagrams (with non-perturbative
propagators) to include in the vertex. At each order, we include the
diagram that recursively dressed the contribution with one less loop
\cite{Bender:2002as,Delbourgo}. However, even at the two loop level,
contributions are already being omitted (compare $\Gamma_{\rm 2\ 
  loop}$ and $\Gamma_{\rm 2\ loop}^\prime$).

With this truncation, Eq.~(\ref{vertex}) reduces to
\begin{equation}
  \Gamma(p,k)= 1 +g^2\int_q D(q) S(p+q) 
  \Gamma(p+q,k+q) S(k+q)\,.
\end{equation}
Using the MN propagator, Eq.~(\ref{MN}), gives the solution
\begin{equation}
  \label{recursive}
  \Gamma_{\rm ladder}=\sum_{i=0}^\infty \left(\alpha\ S^2\right)^i
  =\frac{1}{1 - \alpha\ S^2}\,,
\end{equation}
provided $\alpha\ S^2<1$.  When this is inserted in the propagator DSE
we arrive at a cubic equation for $S$:
\begin{equation}
S^{-1} = x + \frac{\alpha\ S}{1- \alpha\ S^2}\,.
\end{equation}
The physical solution to this equation is given for $\alpha=1$ as the
solid line in Fig.~\ref{fig1}. The sum in Eq.~(\ref{recursive}) is
seen to converge for all space-like and time-like $x$.

\section{Coupled vertex and propagator equations}
\label{fullsolution}

From comparison of the rainbow, one-loop, two-loop ladder
($\Gamma_{\rm 2\ loop}^\prime$), and full ladder solutions of the
previous section (see Fig.~\ref{fig1}), it is apparent that the
approach to this ladder sum is monotonic and inclusion of such ladder
diagrams beyond the first loop only results in small modifications
\cite{Bender:2002as}.  However, further comparison with the full
two-loop vertex solution (corresponding to $\Gamma_{\rm 2\ loop}$)
indicates that the omitted diagrams may be important.  In the simple
MN model defined by Eq.~(\ref{MN}), it is possible to go further and
consistently include all contributions to the vertex. This is the main
result of this article.

The generating functional of the theory we consider is
\begin{eqnarray}
{\cal Z}[J,j]&=& e^{{\cal W}[J,j]}\\
&=&\int {\cal D}\varphi{\cal D}\chi\exp\left[-\int{\cal L}
-(\varphi,J) -(\chi,j)\right]\,,
\nonumber
\end{eqnarray}
where $J$($j$) is an external source for the $\varphi$($\chi$) field,
${\cal L}$ is the sum of the interaction Lagrangian, Eq.
(\ref{lagrangian}), and the usual kinetic pieces and the notation
$(a,b)\equiv\int dx \,a(x)b(x)$. From this we can make a Legendre
transform and define the effective action, ${\cal A}[\varphi,\chi]$
(with $\varphi=-\delta {\cal W}/\delta J$ and
$\chi=-\delta {\cal W}/\delta j$), and thereby the one
particle irreducible vertices,
\begin{equation}
\Gamma^{(n|m)}_{x_1,\ldots,x_n,y_1,\ldots,y_m}=\frac{\delta^{n+m}{\cal
    A}[\varphi,\chi]}{\delta
  \varphi_{x_1}\ldots\delta\varphi_{x_n}\delta
  \chi_{y_1}\ldots\delta\chi_{y_m}}\,.
\end{equation}
The DSEs arise from appropriate functional variation of this effective
action.  As usual, we can identify the position space propagators with
the inverse two-point, one-particle irreducible (1-PI) vertices
\cite{Itzykson:rh}. For example,
\begin{equation}
S(x,y;J)=\frac{\delta^2 {\cal W}}{\delta J_x\delta J_y}
=\left(\frac{\delta^2 {\cal A}}{\delta 
\varphi_x\delta \varphi_y}\right)^{-1}
=\left(\Gamma^{(2|0)}_{x y}\right)^{-1}\,.
\end{equation} 
Thus,
\begin{eqnarray}
K(w,x,y,z)&=&\Gamma^{(4|0)}_{w x y z}
=\frac{\delta^2\Gamma^{(2|0)}_{y z}}{\delta\varphi_w\delta\varphi_x}
\nonumber \\
&=& \frac{\delta \Sigma(y,z;J)}{\delta S(w,x;J)}\,,
\end{eqnarray}
evaluated in the presence of external sources. For the last equality,
we have used the position space analog of the DSE for the $\varphi$
propagator [Eq.~(\ref{phisde})] to express $S^{-1}(x,y;J)$ in terms of
$\Sigma(x,y;J)$, noting that $\delta S_0/\delta S$ vanishes.

By momentum conservation, the full 1-PI four-point scattering kernel,
$K(p,q,r)$, is a function of three external momenta.  However, in the
MN reduction of the theory, no momentum is transferred through $\chi$
propagators and the scattering kernel describes two separate momentum
flows. Consequently we only require $K(p,q,0)$.  Further, since we
only probe the $q=0$ limit of the vertex function in the vertex DSE,
the $\varphi$ propagators in Eq.~(\ref{vertex}) must carry the same
momentum and we only require $K(p,q=p,0)=\delta \Sigma(p)/\delta
S(p)$.

Inserting the self energy from Eq.~(\ref{SE}) and applying the
derivative gives
\begin{widetext}
  \begin{equation}
    \label{kernel}
    K(p,p,0)=\frac{\delta \Sigma(p)}{\delta S(p)}=  \int_q
    \left[\frac{\delta D(q)}{\delta S(p)} S(p-q) \Gamma(p-q,p)+
      D(q)\delta^4(p-q-p)\Gamma(p-q,p) + D(q)S(p-q)\frac{\delta
        \Gamma(p-q,p)}{\delta S(p)} \right] \,.
  \end{equation}
  In principle, Eq.~(\ref{chisde}) determines the functional
  dependence of $D(q)$ on $S(p)$ through the polarisation loop in
  Eq.~(\ref{chiSE}); pre- and post- multiplying both sides of
  Eq.~(\ref{chisde}) by factors of $D(q)$ and taking the derivative
  gives
  \begin{equation}
    \label{ddds}
    -\frac{\delta D(q)}{\delta S(p)}=2 D(q) S(q-p) \Gamma(q-p,p) D(q)
    + D(q) \int_k S(k) \frac{\delta\Gamma(k,k-q)}{\delta S(p)} S(q-k) D(p)\,.
  \end{equation}
\end{widetext}
However, the MN approximation, Eq.~(\ref{MN}), presumably already
invalidates Eq.~(\ref{chisde}) so this result may be misleading. As a
simple alternative to Eq.~(\ref{ddds}), we also consider $\delta
D(p)/\delta S(k)=0$.

We proceed by applying Eq.~(\ref{MN}) to Eqs.~(\ref{SE}),
(\ref{vertex}) and (\ref{kernel}) to give
\begin{equation}
  \label{SEMN}
  \Sigma = \alpha\ S\ \Gamma\,, \\
\end{equation}
\begin{equation}
  \Gamma = 1+ K S^2 \Gamma\,, \\
\end{equation}
\begin{equation}
  K=\alpha \left[\Gamma + S\frac{\delta\Gamma}{\delta S}\right] +K_D\,.
\end{equation}
If Eq.~(\ref{ddds}) is used to determine $\delta D/\delta S$, there is
an additional contribution,
\begin{equation}
  K_D=-\alpha^2 S\ \Gamma\left[2S\ \Gamma + 
    S^2\frac{\delta\Gamma}{\delta S}\right]\,,
\end{equation}
to the kernel. For $\delta D/\delta S=0$, $K_D=0$.  Momentum
dependence is again suppressed in these equations as all quantities
are at the same momentum.  Combining the equations for the vertex and
the kernel gives
\begin{equation}
  \label{vertde}
  \Gamma = 1+ \alpha\ S^2 \Gamma\left(\Gamma\left[1-2 \alpha\ S^2\Gamma\right]
  +S\frac{\delta\Gamma}{\delta S}\left[1-\alpha\ S^2 \Gamma\right]\right)\,,
\end{equation}
for $\delta D/\delta S$ determined from Eq.~(\ref{ddds}), and
\begin{equation}
  \label{vertde0}
  \Gamma = 1+ \alpha\ S^2 \Gamma^2 +\alpha\ S^3
  \Gamma\frac{\delta\Gamma}{\delta S}\,,
\end{equation}
for $\delta D/\delta S=0$.  These functional differential equations
are subject to the physical boundary condition that $\Gamma[0]=1$.
This is simply the requirement that for an infinitely heavy $\varphi$
particle (whose propagator $S=1/m^2\stackrel{m\to\infty}
{\longrightarrow} 0$), the vertex reduces to the bare vertex as loop
corrections involving the $\varphi$ propagator vanish.  Either of
these equations, together with their boundary condition\footnote{At
  $x=0$, the propagator DSE reduces to $\Gamma[S]=1/(\alpha S^2)$
  which, when inserted in the vertex equation [Eq.~(\ref{vertde0})],
  implies $S=\pm\sqrt{2/\alpha}$ and $\Gamma[S]=\frac{1}{2}$.
  However, here the functional dependence of the vertex is
  incompatible with the boundary condition and the solution is
  unphysical.} and Eqs.~(\ref{phisde}) and (\ref{SEMN}) close provided
the functional derivative $\delta\Gamma/\delta S$ can be
determined.\footnote{A related method of functional closure was
  presented very recently in a perturbative context in
  Ref.~\cite{Pelster:2001cc}.}

\subsection{Diagram counting}
\label{diagrams}

Physically, the use of the Munczek-Nemirovsky-like propagator,
Eq.~(\ref{MN}), means that $\chi$ field loops dressing the propagation
of the $\varphi$ field or dressing the vertex carry no momentum.
Consequently, objects such as the $\varphi$ self energy and the vertex
can be expressed as appropriately weighted sums of products of dressed
$\varphi$ propagators at the same momentum. That is, $\Gamma$ and
$\Sigma$ are analytic functionals of $S$ and can be written as
\begin{equation}
  \label{vertexsum}
  \Gamma = \sum_{i=0}^\infty f_i(\alpha\ S^2)^{i}\,,
\end{equation}
and
\begin{equation}
  \Sigma = \alpha\ S \sum_{i=0}^{\infty}e_i(\alpha\ S^2)^{i}\,.
\end{equation}
As a corollary, $\delta\Gamma/\delta S$ exists and can be
defined. Indeed Eqs.~(\ref{vertde}) and (\ref{vertde0}) can both be
rewritten as recursions amongst the coefficients $f_i$. For $\delta
D/\delta S=0$, the relation is
\begin{equation}
  \label{coefficients}
  f_i=\delta_{i0}+\sum_{j=0}^{i-1}(2j+1)f_j f_{i-j-1}\,. 
\end{equation}
These coefficients simply count the number of 1-PI contributions to
the vertex dressing with $i$ loops in terms of full propagator
lines.\footnote{The enumeration of this series is considered in
  Refs.~\cite{Petermann,Cvitanovic:1978wc,Stein,Broadhurst:1999ys},
  but no closed form for the $i^{th}$ element exists to the author's
  knowledge and the recursion of Eq.~(\ref{coefficients}) provides an
  efficient calculational method.} The first few coefficients are: 1,
1, 4, 27, 248, 2830, \ldots\ and their asymptotic behaviour is $f_n\sim
(2n+1)!!/e$ \cite{Stein} from which it is apparent that the sum in
Eq.~(\ref{vertexsum}) does not converge (nor is it Borel summable) for
$\alpha\ S^2 >0$.  For the case of $\delta D/\delta S$ given by
Eq.~(\ref{ddds}), an additional term
\begin{displaymath}
  -2\sum_{j=0}^{i-2}\sum_{k=0}^{i-j-2}(k+1)f_j f_k f_{i-j-k-2}
\end{displaymath}
appears on the right-hand side of Eq.~(\ref{coefficients}) and the
series is similarly divergent.  The self energy coefficients, $e_i$,
can also be calculated and a generating function can be constructed
for them \cite{Arques}.

\begin{figure}
  \includegraphics[width=0.95\columnwidth]{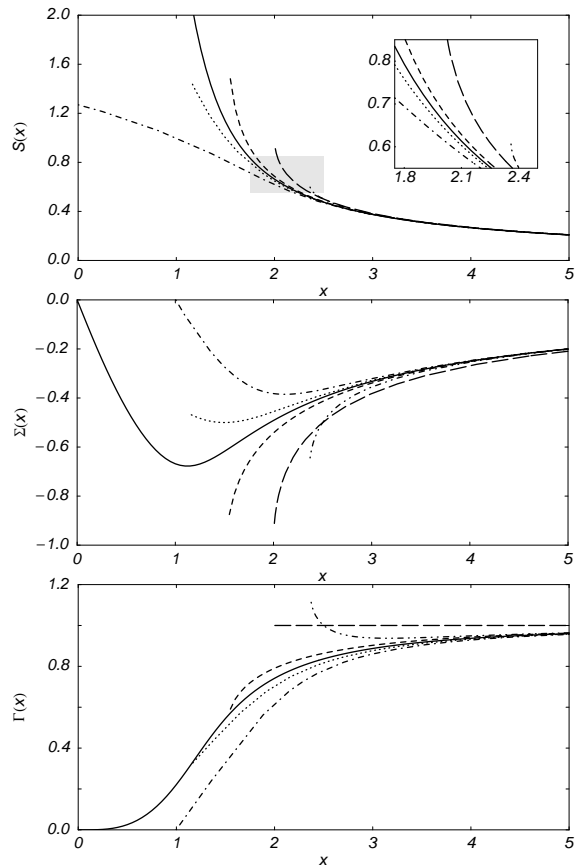}
  \caption{\label{fig2} Dependence of the propagator (upper
    panel), self-energy (middle panel) and vertex function (lower
    panel) on the variable $x=p^2+m^2$ for $\alpha=-1$. Solutions are
    shown for the bare- (long-dashed), the one-loop- (dot-dashed),
    two-loop- (dot-dot-dashed) vertices, for the ladder truncation of
    the vertex DS equation (dotted), for the full solution with
    $\delta D/\delta S=0$ (short-dashed) and for the full solution
    with $\delta D/\delta S$ from Eq.~(\ref{ddds}) (solid). To clarify
    the various solutions for the propagator, the shaded region is
    expanded in the inset.}
\end{figure} 
The apparent lack of solution is supported by the failure of various
standard numerical methods \cite{NumRec} to find a solution to either
of the differential equations for $\Gamma$ with $\alpha>0$. This
result is perhaps not unexpected since the theories defined by the
Lagrangian of Eq.~(\ref{lagrangian}) and the related $\varphi^3$
theory are unstable as they have no ground state
\cite{Dyson:tj,Baym,Tjon,Schreiber,Cornwall:1995dr}. However, the
effect that the use of the MN model for the $\chi$ propagator has on
these conclusions is difficult to assess.

\subsection{Numerical solution for $\alpha<0$}

If we consider $\alpha<0$ (as discussed at the beginning of
Sec.~\ref{truncate}, the magnitude of $\alpha$ is unimportant),
solutions of Eqs.~(\ref{vertde} )and (\ref{vertde0}) exist. For this
to be the case, the theory either has an imaginary coupling, $g=i h$
($h\in R\!\!\!\!\!\!I\;$), or the correlation between the $\chi$
fields at two points is negative so the fields are imaginary valued.
The first scenario corresponds to a non-Hermitian Hamiltonian that is
${\cal PT}$-symmetric. Such theories retain a positive energy spectrum
and have been considered in detail by Bender {\it et al.}
\cite{Bender}.

The numerical solution of the vertex differential equations and their
boundary condition is straightforward for $\alpha<0$ (the functional
nature is irrelevant because of the decoupling of propagators at
different momenta). Once we have determined the functional dependence
of the vertex on $S$, we can combine that with Eqs.~(\ref{phisde}) and
(\ref{SEMN}) and obtain exact solutions to the MN model.

Figure~\ref{fig2} depicts the momentum dependence of the $\varphi$
propagator, the self-energy and the vertex as a function of $x$ for
$\alpha=-1$ and for the two alternate determinations of $\delta
D/\delta S$. The solid line corresponds to $\delta D/\delta S$
determined from Eq.~(\ref{ddds}), while the short-dashed line is for
$\delta D/\delta S=0$. For comparison, we also show the ($\alpha=-1$)
solutions obtained using various of the truncations of the preceding
section. For $\alpha<0$, most of the algebraic solutions become
complex at some critical value of $x$, hence the abrupt termination of
the solutions in the figure.  Numerically we find that the full
solution for $\delta D/\delta S=0$ has a critical point $x_{\rm
  crit}=1.518\sqrt{|\alpha|}$, although it is not known whether this
critical behaviour has any physical significance. For $\delta D/\delta
S$ determined from Eq.~(\ref{ddds}), the solution exists for all $x$.
It is also evident from the figure that the various analytic solutions
beyond the bare vertex provide a reasonable approximation to the full
solution for moderate to large momenta as should be expected.

\subsection{Higher-point vertices}

\begin{figure}[!t]
  \includegraphics[width=0.95\columnwidth]{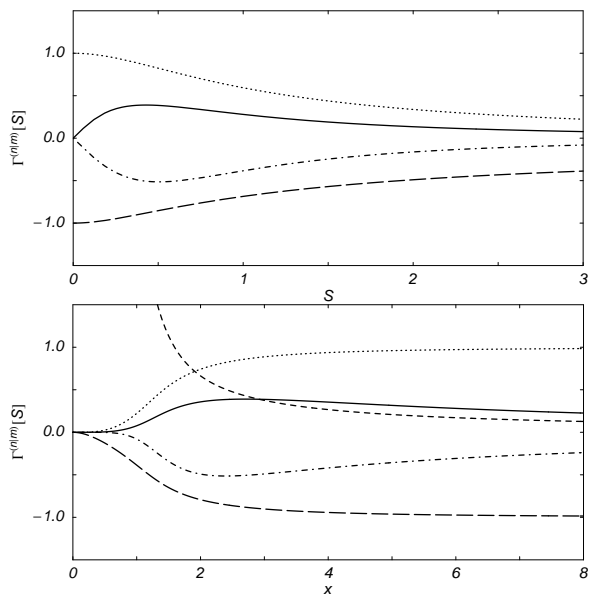}
  \caption{\label{higher} Behaviour of the higher point Green functions as
    functionals of $S$ and functions of $x$ as described in the text.}
\end{figure}
We now know the functional dependence of $\Gamma[S]$
($=\Gamma^{(2|1)}$) and, through Eq.~(\ref{kernel}), of $K[S]$
($=\Gamma^{(4|0)}$) for $\alpha<0$. Consequently, we can (numerically)
construct every 1-PI Green function of the theory with zero or one
external $\chi$ leg carrying zero momentum and any number of pairs of
$\varphi$ legs carrying the same momentum.  Figure~\ref{higher} shows
$\Gamma^{(2|0)}$ (short-dashed), $\Gamma^{(2|1)}$ (dotted),
$\Gamma^{(4|0)}$ (long-dashed), $\Gamma^{(4|1)}$ (dot-dashed),
$\Gamma^{(6|0)}$ (solid) as functionals of $S$ (upper panel) and as
functions of $x$ (lower panel) given the numerical solution for $S$,
all with $\alpha=-1$ and $\delta D/\delta S$ obtained from
Eq.~(\ref{ddds}).

Since we do not know the behaviour of $K(p,q\ne p,0)$, we cannot apply
the results of this analysis to the study of massive bound states
within the model as was done in Ref.~\cite{Bender:2002as} for the
ladder vertex truncation.

\section{Summary}
\label{conclusion}

The analysis in this paper demonstrates that the exact MN model
defined by Eq.~(\ref{MN}) has no solutions for positive values of the
coupling $\alpha$. However when additional truncations are introduced
for the vertex or for the scattering kernel, solutions can be found.
(It is of course perfectly reasonable to {\it define} the MN model to
include the bare vertex truncation.) For the case of $\alpha<0$ we
solve the full MN model and investigate how well various truncations
approximate the full solution.

The extension to a more QCD-like theory (which was the motivation for
this study) through the use of fermion and vector boson fields appears
possible after the complications introduced by the Dirac structure are
addressed. This will result in coupled, partial differential equations
for the propagator dependence of the various Dirac components of the
fermion-gauge boson vertex.  Since QCD and more realistic models of it
do not suffer from the fundamental difficulties of the scalar theory
considered here, the problems of solution existence that we
encountered would likely not persist.  As a further extension of this
work, there may be some scope to extend this analysis to models with
less restrictive $\chi$ propagators such as Gaussian forms
\cite{gaussian}.

\acknowledgments The author is appreciative of numerous discussions
with A.~Schreiber and comments from R.~Alkofer, M.~Oettel,
C.~D.~Roberts and M.~Savage. This work was supported by the Centre for
the Subatomic Structure of Matter at the University of Adelaide, DFG
grant Al279/3-3, and DOE grant DE-FG03-97ER41014.

\bibliographystyle{h-physrev}

\end{document}